\documentclass[aps,prb,preprint,unsortedaddress,showkeys]{revtex4-2}

\usepackage{graphicx}
\usepackage{amsmath}

\begin{document}

\title{Optically Induced Avoided Crossing in Graphene}

\author{Sören Buchenau}
\email[]{sbuchena@physnet.uni-hamburg.de}
\author{Benjamin Grimm-Lebsanft}

\author{Florian Biebl}

\author{Tomke Glier}

\author{Lea Westphal}

\affiliation{Institute of Nanostructure and Solid State Physics, University of Hamburg, 22761, Germany}

\author{Janika Reichstetter}
\author{Dirk Manske}
\affiliation{Max Planck Institute for Solid State Research, 70569 Stuttgart, Germany}

\author{Michael Fechner}
\author{Andrea Cavalleri}
\affiliation{Max Planck Institute for the Structure and Dynamics of Matter, 22761 Hamburg, Germany}

\author{Sonja Herres-Pawlis}
\affiliation{Institute of Inorganic Chemistry, RWTH Aachen University, 52074 Aachen, Germany}

\author{Michael Rübhausen}
\email[]{mruebhau@physnet.uni-hamburg.de}
\affiliation{Institute of Nanostructure and Solid State Physics, University of Hamburg, 22761, Germany}

\date{\today}

\begin{abstract}
Degenerate states in condensed matter are frequently the cause of unwanted fluctuations, which prevent the formation of ordered phases and reduce their functionalities. Removing these degeneracies has been a common theme in materials design, pursued for example by strain engineering at interfaces. Here, we explore a non-equilibrium approach to lift degeneracies in solids. We show that coherent driving of the crystal lattice in bi- and multilayer graphene, boosts the coupling between two doubly-degenerate modes of E\textsubscript{1u} and E\textsubscript{2g} symmetry, which are virtually uncoupled at equilibrium. New vibronic states result from anharmonic driving of the E\textsubscript{1u} mode to large amplitdues, boosting its coupling to the E\textsubscript{2g} mode. The vibrational structure of the driven state is probed with time-resolved Raman scattering, which reveals laser-field dependent mode splitting and enhanced lifetimes. We expect this phenomenon to be generally observable in many materials systems, affecting the non-equilibrium emergent phases in matter. 
\end{abstract}

\keywords{Phononics, Raman, Pump-Probe, Graphene}

\maketitle

\section{\label{sec:introduction}Introduction}
In condensed matter systems and at interfaces, static symmetry breaking has been used to create new states of matter and to introduce new functionalities in equilibrium \cite{von1979devil,hasan2010colloquium,dong2015multiferroic}. However, using coherent and dynamic optical control offers an innovative approach to expand this concept beyond the limitations of static control \cite{juraschek2017dynamical,mciver2020light} and on short or even ultrashort time scales. There are many examples where coherent dynamical control of matter by photons has triggered phase transitions, tuned interactions, and generated new forms of matter \cite{Bloch2022}. Even though some of these states exist only for picoseconds, it is possible to investigate them using pulsed lasers and pulsed currents in transport measurement \cite{Wessels2016,Disa2020}. The structural degrees of freedom, represented by phonon modes, are essential for the thermodynamic properties in crystals. Thus, materials with degenerate phonons can exhibit strong anharmonic couplings leading to new vibronic states resulting in new forms of matter.

Graphene systems have been previously excited mostly by high-photon-energy excitations, in resonance with electronic transitions over a broader energy range of 1-2 eV \cite{yan2009time,kang2010lifetimes}. Here, however, we drive structural degrees of freedom in resonance with a low-energy degenerate IR phonon of bilayer graphene. The E\textsubscript{1u} IR phonon and E\textsubscript{2g} Raman active phonon are uncoupled in equilibrium. We resonantly drive the infrared vibration at 196~meV ($\sim$~48 THZ) using a mid-IR pump. The vibronic state of the driven system is then monitored via the corresponding transient Raman response. We explore the physics of lifting degeneracies by optical control in bi- and multilayer graphene and demonstrate the emergence of an optically induced avoided crossing that can be tuned by the fluence of the pump laser. 

\begin{figure}
\includegraphics{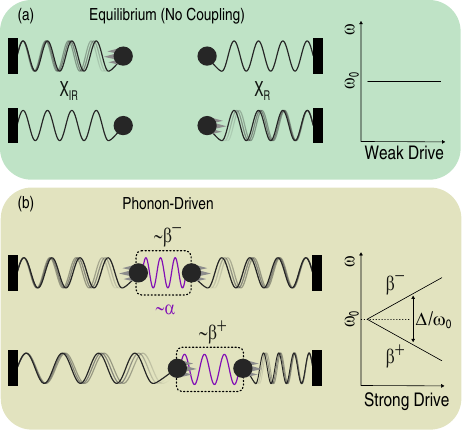}
\caption{\label{fig:figure1}Optically induced avoided crossing shown with classical springs and masses. (a) The IR oscillator X\textsubscript{IR} and the Raman active oscillator X\textsubscript{R} are degenerate at $\omega_0$, but exhibit different symmetries, i.e., odd and even. Therefore, without any coupling, an oscillation of either mode does not impact the other one. (b) When the IR mode is strongly driven out of the harmonic approximation, the anharmonic coupling $\alpha$ couples the oscillators and the vibronic systems form the two novel vibrational states $\beta^-$ and $\beta^+$. The vibration $\beta^-$ is antisymmetric, i.e. the masses move in opposite directions, whereas the vibration $\beta^+$ is symmetric, i.e. the masses move in the same direction. The introduced energy difference between the two states, $\Delta$, is given by the fluence of the pump laser. The Raman-active superposition $\beta^+ - \beta^-$ can be experimentally measured and exhibits a subharmonic frequency, tunable by the field-dependent splitting $\Delta$.}
\end{figure}

Monolayer graphene exhibits the prominent G-phonon \cite{malard2009raman,nemanich1977infrared}, which is only Raman-active. On the other hand, from two layers and up, the material exhibits fully degenerate IR- and Raman-active phonons with orthogonal E\textsubscript{1u} and E\textsubscript{2g} symmetry \cite{nemanich1977infrared,ferrari2013raman,davydov1964theory}. These modes are ideal candidates for strong anharmonic coupling due to their strongly related eigenvectors and degenerate energies. Figure~\ref{fig:figure1}\,(a) shows the IR- and Raman-active oscillators on each side. Without coupling in equlibrium, the phonons are at the same eigenfrequency. Figure~\ref{fig:figure1}\,(b), driving the IR mode leads to a lifting of the degeneracy via anharmonic coupling of the two phonon modes \cite{forst2011nonlinear}, which is indicated in the figure symbolically as an additional spring coupling the IR and Raman active phonons. This optically induced anharmonic coupling out of equilibrium lifts the degeneracy resulting in two new vibrational states $\beta^+$ and $\beta^-$, representing two coupled harmonic oscillators. The vibration $\beta^-$ is antisymmetric, i.e., the masses move in opposite directions, whereas the vibration $\beta^+$ is symmetric, i.e. the masses move in the same direction. From those, the IR- and Raman-active modes can be obtained by $\beta^+ +\beta^-$ and $\beta^+ -\beta^-$, respectively. Hence the transient Raman signal vanishes in the degenerate state, i.e. $\beta^+ = \beta^-$ and is, therefore, a precise tool to study the anharmonic coupling out of equilibrium.

If one considers a relative shift of the respective eigenvalues in the limit of small Rabi frequencies, the energy-shifted values $\beta^-$ and $\beta^+$ can be expressed by $\omega_\pm\approx\omega_0\left(1\pm\Delta/2\omega_0^2\right)$, with $\omega_0$ as the degenerate frequency of the phonons in the equilibrium state and $\Delta$ the representing lifting of the degeneracy due to the anharmonic coupling out of equilibrium. Thus, the orthogonal harmonic motions of the IR- and Raman-active vibrations are $\sim \text{e}^{i \omega_0 t} \sin{\left( \Delta \cdot t/2\omega_0 \right)}$ and $\sim \text{e}^{i\omega_0 t}\cos{\left(\Delta\cdot t/2\omega_0 \right)}$, respectively. Accordingly, the IR and Raman coordinates vibrate with the original degenerate eigenfrequency $\omega_0$, modulated by a subharmonic frequency and are phase-shifted by $\pi/2$. In static quantum systems, the avoided crossing \cite{landau1932theory,zener1932non} phenomenon leads to lower eigenvalues, which increases the stability of bonded states in molecular systems. In optically driven non-equilibrium systems, the magnitude of degeneracy lifting $\Delta$ depends on the strength of the electric field as it tunes the anharmonic coupling. As a result, the populations of the lower and higher energy states undergo time-dependent oscillations. The transient spontaneous Raman scattering of a Raman-active phonon, driven by an IR-active phonon, is the ideal method to observe these states. However, this observation is only possible in case of a symmetry breaking yielding a finite response due to an optically driven avoided crossing, i.e., when $\beta^+\neq\beta^-$.

\section{\label{sec:experiment}Experiment}
\begin{figure}
\includegraphics{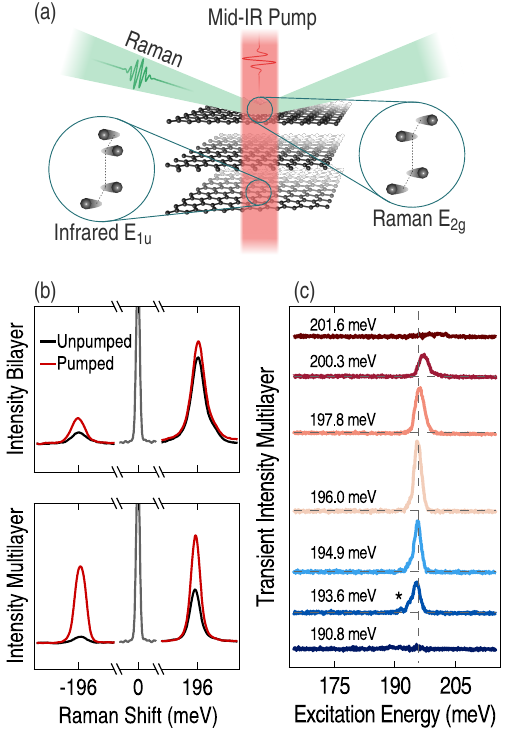}
\caption{\label{fig:figure2}Outline of the transient spontaneous Anti-Stokes and Stokes Raman scattering experiment. (a) A mid-IR pump pulse excites the IR vibration in the sample, which is probed by a Raman pulse. The delay, pump energy, and fluence are controlled. The insets show the movements of the carbon atoms during the vibrations. (b) Resonant mid-IR pumping causes a strong enhancement of the signal in bi- and multilayer graphene. (c) Selective driving of the E\textsubscript{1u} vibration by tuning the pump energy through the sharp resonance. Note that small visible features (see asterisk) might be due to sum-frequency scattering. Horizontal grey dashed lines indicate the zero line for each spectrum and vertical ones the energy of the unpumped IR phonon.}
\end{figure}

The general outline of the experiment is shown in Figure~\ref{fig:figure2}\,(a). A tunable mid-IR pulse drives the E\textsubscript{1u} IR phonon mode in resonance at 196\,meV. At different time delays, transient spontaneous Raman scattering probes the Raman-active E\textsubscript{2g} mode. In bi- and multilayer systems, the presence of the polar IR-active phonon results in a strong transient response in the excited state, as seen in Figure~\ref{fig:figure2}\,(b). The transients are obtained by subtracting the unpumped signal from the pumped signal, isolating the response of the IR-driven Raman mode. Note that monolayer graphene lacks the polar vibrational mode \cite{malard2009raman} and, therefore, does not exhibit a transient response (see SI Appendix, Figure S2). In multilayer graphene, the spontaneous Raman signal in the pumped state is enhanced on the Stokes and Anti-Stokes side.

In the following, we show the transient spontaneous Raman data on the Anti-Strokes side with the expected stronger transient as the pump excites the coupled vibronic system out of equilibrium. The transient Anti-Stokes Raman response as a function of the energy of the mid-IR pump is shown in Figure~\ref{fig:figure2}\,(c). The first apparent observation is the ultra-sharp resonance when tuning the mid-IR laser through the energy of the polar phonon. We observe a peaking at 196\,meV with a width of about 6\,meV. This observation, together with the absence of a transient Raman response in monolayer graphene (see SI Appendix, Figure S2) and the absence of any transients above and below the energy of the polar IR phonon (see SI Appendix, Figure S4 (a) and Figure~\ref{fig:figure2}\,(c), highlights the direct coupling of the transient Raman E\textsubscript{2g} phonon to the IR-active E\textsubscript{1u} phonon. Note that we do not observe any electronic background, as might be expected when pumping at optical energies (see SI Appendix, Figure S4 (b)). Additionally, we observe changes in the shape of the transient Anti-Stokes Raman response as the energy of the mid-IR pump is changed and slightly detuned from the energy of the E\textsubscript{1u} phonon. Thus, the data show a purely phononic mechanism for driving graphene systems.

The incident mid-IR pump photon drives the E\textsubscript{1u} phonon, which alters the response of the Raman signal of the E\textsubscript{2g} phonon. The degeneracy of the two phonons leads to strong anharmonic coupling \cite{forst2011nonlinear,nicoletti2016nonlinear}. Without the pump interaction, the Raman response is given by the Raman susceptibility, yielding a typical Lorentzian lineshape. The anharmonic coupling $\alpha$ between a Raman and IR-active phonon requires a quadratic coupling to the Raman-active phonon in the equation of motion as a consequence of a cubic contribution to the potential, i.e. the anharmonic coupling. This leads to a set of equations of motions describing the electric field-driven IR-active phonon $X_{\text{IR}}(t)$ and Raman-active phonon $X_{\text{R}}(t)$

\begin{subequations}
\begin{align}\label{equ:equ1}
\widetilde{D}(\omega_{0},\gamma_{0}) X_{\text{IR}}(t) &= \alpha X_{\text{R}}(t) X_{\text{IR}}(t) + F_{\text{L}}(t)\\
\widetilde{D}(\omega_{0},\gamma_{0}) X_{\text{R}}(t) &= \alpha X^2_{\text{IR}}(t).
\end{align}
\end{subequations}

Here, $\widetilde{D} \left( \omega_0, \gamma_0 \right) = \partial_t^2+\gamma_0\partial_t+\omega_0^2$ is the operator describing a damped harmonic oscillator and $F_\mathrm{L}\left(t\right)$ represents the Gaussian laser beam with frequency $\omega_\mathrm{L}$ and width $\sigma$. Assuming that IR- and Raman-active phonons are degenerate, the above set of coupled nonlinear differential equations can be solved analytically. For this procedure, one adds and subtracts both equations from each other and introduces common coordinates, linking the coordinates of the IR- and Raman-active phonons by $\beta^+ = X_{\mathrm{IR}} + X_\mathrm{R}$ and $\beta^-=X_{\mathrm{IR}}-X_\mathrm{R}$. The inverse Fourier transformation of $\beta^\pm$ then yields the expressions for these effective phonon propagators after applying the convolution theorem for the quadratic parts of the equation

\begin{subequations}
\begin{align}
\beta^+ (\omega) &= \frac{\alpha \left[(\beta^- * \beta^+) (\omega) \right]+ 2 F_\text{L}(\omega)}{2(\omega_{-}^2 - \omega^2 + i \gamma_{-} \omega)} \label{equ:equ2a}\\[1.5ex]
\beta^- (\omega) &= \frac{ -\alpha \left[(\beta^+ * \beta^- ) (\omega) \right]  + 2 F_\text{L}(\omega)}{2(\omega_{+}^2 - \omega^2 + i \gamma_{+} \omega)},\label{equ:equ2b}
\end{align}
\end{subequations}

with the shifted eigenfrequencies $\omega_\pm^2=\omega_0^2\pm\Delta$ and modified damping constants $\gamma_\pm=\gamma_0\pm\delta/\omega$. These changes scale with the anharmonic coupling constant $\alpha$ and the square of the electric field $E_0^2$. The modifications $\Delta$ and $\delta$ represent the frequency-dependent contributions from the real and imaginary parts of the convolution integrals. The contribution from the non-mixed convolution integrals $\left(\beta^+\ast\beta^+\right)\left(\omega\right)$ and $\left(\beta^-\ast\beta^-\right)\left(\omega\right)$ vanishes and they are omitted in eq. (\ref{equ:equ2a}) and (\ref{equ:equ2b}). The mixed terms $\left(\beta^-\ast\beta^+\right)\left(\omega\right)$ have to be evaluated self-consistently. In the following, we assume $\Delta$ and $\delta$ to be frequency independent parameters to model the Raman response. However, both parameters depend on time and the fluence of the pump $F_\mathrm{L}$. The leading order contribution results in the modified propagators with the driving pump laser in the numerators, shown in eq. (\ref{equ:equ2a}) and (\ref{equ:equ2b}). The Green's function of the Raman-active vibration can now be calculated by $X_\mathrm{R}\left(\omega\right)=\frac{1}{2}\left(\beta^+-\beta^-\right)$ and the Raman scattering intensity is given by $-\mathrm{Im}\left[X_\mathrm{R}\left(\omega\right)\right]$

\begin{subequations}
\begin{align}\label{equ:equ3a}  \nonumber
I(&\omega) = \frac{ F_\text{L}(\omega)  \Delta \omega (\gamma_+(\omega_-^2\!-\!\omega^2)\!+\!\gamma_-(\omega_+^2\!-\!\omega^2)) }{\mathcal{D}}  + \\[1.5ex]
& \frac{ F_\text{L}(\omega) \delta (\gamma_- \gamma_+ \omega^2 \!-\! (\omega_-^2\!-\!\omega^2) (\omega_+^2\!-\!\omega^2) ) }{\mathcal{D}}
\end{align}
\begin{align}
\begin{split}\label{equ:equ3b}
\mathcal{D}=((\omega^2_+-\omega^2)(\omega^2_--\omega^2) - \gamma_- \gamma_+ \omega^2)^2\\[1.5ex]
+ \omega^2(\gamma_+(\omega_-^2-\omega^2) + \gamma_-(\omega_+^2-\omega^2))^2
\end{split}
\end{align}
\begin{align}
\begin{split}\label{equ:equ3c}
\hspace{-0.15cm}F_\text{L}(\omega)\!=\!\frac{f_\text{L}}{\sqrt{2 \pi}} (e^{-\frac{1}{2} \sigma^2 (\omega - \omega_\text{L})^2}\!+\!e^{-\frac{1}{2} \sigma^2 (\omega + \omega_\text{L})^2}).
\end{split}
\end{align}
\end{subequations}

\begin{figure*}
\includegraphics{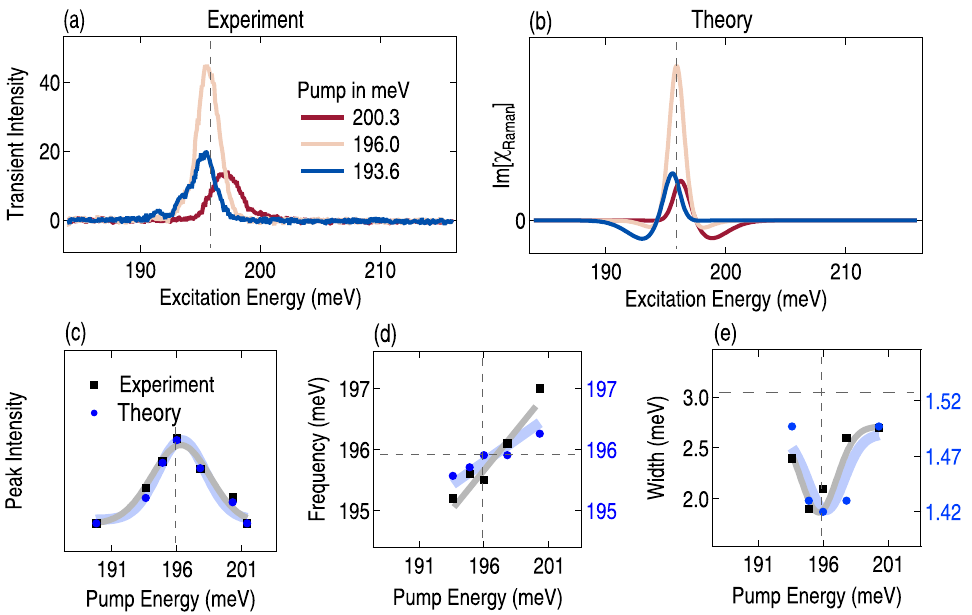}
\caption{\label{fig:figure3}Experimental and calculated transients for mid-IR resonant driving. (a) Exemplary transient spectra around the resonance energy of the E\textsubscript{1u} mode for multilayer graphene. All non-trivial transients are shown in Figure~\ref{fig:figure2}\,(c). Note that small visible features (see asterisk) might be due to sum-frequency scattering. (b) Simulation results obtained by eq. (\ref{equ:equ3a}) for the values of $\Delta= 4$\,meV$^2$ and $\delta=0.12$\,meV$^2$. (c) Intensity scaling of the transient as a function of pump energy. (d) Frequency tuning of the transient as a function of pump energy. (e) Change of the width as a function of pump energy. In (c) to (e) black squares represent experimental results and blue circles the calculation. Horizontal grey dashed lines indicate the frequency and width of the unpumped Raman phonon and vertical ones the energy of the IR phonon. Solid grey and blue lines are guides to the eye for each parameter.}
\end{figure*}

As expected, the transient Raman response vanishes for $\omega_+=\omega_-=\omega_0$, i.e. $\Delta \rightarrow 0$  and $\gamma_+=\gamma_-=\gamma_0$, i.e., $\delta\rightarrow 0$, representing the case of two non-interacting oscillators at identical frequencies and $\alpha=0$. The response from the above expression exhibits a classical anti-resonance contribution in the first term and a sharpened Lorentzian contribution from the second term. It is, therefore, the unusual case of a Fano-like response resulting from the interaction of two discrete states, which is quite different from the typical case of a Fano lineshape where a discrete excitation interacts with a continuum of states as in electron-phonon coupled systems \cite{fano1961effects,hasdeo2014breit,zhang2015observation,bock1999anomalous}. It is also remarkable that the strength of the change in frequency and width, i.e., the changes in real and imaginary parts of the phonon susceptibility, directly influences the lineshape and the strength of the response. Both are proportional to the square of the electric field, which drives the IR phonon. Stronger changes in the frequency lead to a more asymmetric Fano-like lineshape, and larger changes in the width lead to a more symmetric Lorentzian response. We can now model the observed resonance interaction between IR- and Raman-active phonons, as shown in Figure~\ref{fig:figure3}. The frequencies and widths $\omega_0$, $\gamma_0$, $\omega_L$, and $\sigma$ are known. The only two model parameters are the modifications to the frequency $\Delta$ and the width $\delta$ as outlined above. The lineshape obtained is only dependent on the ratio of the parameters $\Delta$ and $\delta$, so these values are subject to an unknown scaling parameter that is given by the enhancement of the response. The simulation results are shown in Figure~\ref{fig:figure3}\,(a) and (b) for the experimental and calculated transients. The data are reproduced very well, considering that the frequency-dependent quantities $\Delta$ and $\delta$, are modeled by constants. The signal enhancement, the shifting of mode frequencies, and the sharpening of the transient response are recovered by the simulation where the ratio of $\Delta/\delta$ was kept constant and set to $\Delta=4$\,meV$^2$ and $\delta=0.12$\,meV$^2$. The two anti-resonances in the transient in Figure~\ref{fig:figure3}\,(b) are not observed in the experiment. We attribute these differences to higher-order coupling terms that we have neglected in our calculation and to the fact that $\Delta$ and $\delta$ should exhibit a frequency dependence. The shape of the resonance tuning of the intensity and the absolute frequency shifts are well recovered, as can be seen in Figure~\ref{fig:figure3}\,(c) and (d). Even the effect of the sharpening as such is present in our calculation in Figure~\ref{fig:figure3}\,(e). However, the details of the broadening when tuning out of the resonance are not captured. The FWHM of the Raman mode measured in the probe spectra is 3\,meV, the width of the observed transient in resonance is 2\,meV, and the width of the calculated transient is 1.4\,meV. The linewidth variation with the pump energy is well reproduced but takes place on a smaller scale than the experiment.

Additionally, assuming a time-dependent dephasing between the IR- and Raman-active modes, the ratio between $\Delta$ and $\delta$ in eq. (\ref{equ:equ3a}) will increase with time. The calculation recovers this delay dependence between pump and probe pulses, as shown in the SI Appendix Figure S3, demonstrating a transient state with a lifetime of several picoseconds. This lifetime is remarkable since, in the static limit, the widths of the phonons correspond to a lifetime of about 200\,fs. The observation of the transient Raman response on the time scale of several picoseconds implies an increased robustness of the non-equilibrium coupled driven state compared to the equilibrium. This is further supported by the fact that the transient Raman response is substantially sharper than its static spontaneous Raman response.

\begin{figure*}
\includegraphics{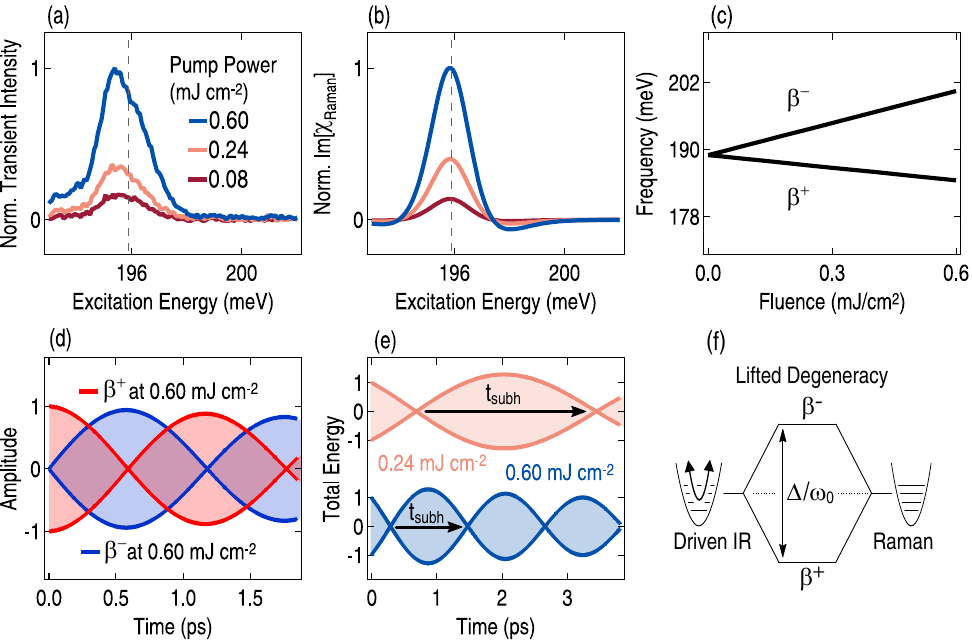}
\caption{\label{fig:figure4}Fluence-dependence of the lifting of the degeneracy. (a) Fluence-dependent transient signals. (b) Calculated transient signals obtained by equation (3a). (c) DFT calculation of the splitting of the frequencies as a function of the fluence of the driving IR field. (d) Amplitude of the $\beta^+$ and $\beta^-$ oscillations at 0.60\,mJcm$^{-2}$. The data are derived from the simulation and by comparison to the DFT calculated fluence-dependent splitting of the frequencies. (e) Total energy from the mixed vibrational state for 0.24 and 0.60\,mJcm$^{-2}$. The magnitude t\textsubscript{subh} scales with the lifting of the degeneracy $\Delta$. (f) Schematic of the field-dependent splitting of the oscillations.}
\end{figure*}

Increasing the fluence at constant mid-IR pump frequency and delay leads to a linearly increasing transient signal, as shown in Figure~\ref{fig:figure4}\,(a), which is well recovered by the simulation in Figure~\ref{fig:figure4}\,(b). The symmetry breaking of the degenerate energy levels is directly proportional to the fluence of the mid-IR pulse, and so is the transient Raman signal. Figure~\ref{fig:figure4}\,(c) shows a DFT calculation of the magnitude of the energy splitting of the common coordinates $\beta^+$ and $\beta^-$ as a function of the field strength of the driving mid-IR field. The DFT calculation allows now to determine real values of $\Delta$ and $\delta$, and to calculate the effective vibrational patterns as shown in Figure~\ref{fig:figure4}\,(d), clearly resembling a novel non-equilibrium system of two coupled quantum states. Figure~\ref{fig:figure4}\,(e) shows the total energy of the 0.24\,mJcm$^{-2}$ and 0.60\,mJcm$^{-2}$ fluence data. It shows the period of the subharmonic Rabi oscillation t\textsubscript{subh}, which depends on the fluence. Accordingly, it is inversely proportional to the fluence-dependent splitting of the avoided crossing $\Delta$, as shown in Figure~\ref{fig:figure4}\,(f). The splitting separates the antisymmetric and symmetric vibrations $\beta^+$ and $\beta^-$ as shown in Figure~\ref{fig:figure1}\,(b).

In summary, we have shown that degenerate phonons in bi- and multilayer graphene can be used to generate novel vibronic states due to an induced anharmonic coupling, tuned by driving a polar phonon mode. Coherent control over the vibronic states can be obtained by tuning the electric field of the IR laser driving the phonon. The lower and higher energy states correspond to novel symmetric and antisymmetric vibronic states, resembling the physics of the avoided crossing. These states exchange population density at a subharmonic Rabi frequency, defined by the applied electric field, effectively leading to a breaking of continuous time-translation symmetry. The theoretical as well as experimental framework can be extended towards other 2D materials with degenerate or nearly degenerate phonon modes evolving the field of nonlinear phononics. 
\begin{acknowledgments}
We acknowledge funding by the Deutsche Forschungsgemeinschaft via RU 773/8-1 and the Bundesministerium für Bildung und Forschung via 05K19GU5 and 05K22GU2.
\end{acknowledgments}

\bibliography{Optically_Induced_Avoided_Crossing}

\providecommand{\noopsort}[1]{}\providecommand{\singleletter}[1]{#1}%
\begin{thebibliography}{22}%
\makeatletter
\providecommand \@ifxundefined [1]{%
 \@ifx{#1\undefined}
}%
\providecommand \@ifnum [1]{%
 \ifnum #1\expandafter \@firstoftwo
 \else \expandafter \@secondoftwo
 \fi
}%
\providecommand \@ifx [1]{%
 \ifx #1\expandafter \@firstoftwo
 \else \expandafter \@secondoftwo
 \fi
}%
\providecommand \natexlab [1]{#1}%
\providecommand \enquote  [1]{``#1''}%
\providecommand \bibnamefont  [1]{#1}%
\providecommand \bibfnamefont [1]{#1}%
\providecommand \citenamefont [1]{#1}%
\providecommand \href@noop [0]{\@secondoftwo}%
\providecommand \href [0]{\begingroup \@sanitize@url \@href}%
\providecommand \@href[1]{\@@startlink{#1}\@@href}%
\providecommand \@@href[1]{\endgroup#1\@@endlink}%
\providecommand \@sanitize@url [0]{\catcode `\\12\catcode `\$12\catcode
  `\&12\catcode `\#12\catcode `\^12\catcode `\_12\catcode `\%12\relax}%
\providecommand \@@startlink[1]{}%
\providecommand \@@endlink[0]{}%
\providecommand \url  [0]{\begingroup\@sanitize@url \@url }%
\providecommand \@url [1]{\endgroup\@href {#1}{\urlprefix }}%
\providecommand \urlprefix  [0]{URL }%
\providecommand \Eprint [0]{\href }%
\providecommand \doibase [0]{https://doi.org/}%
\providecommand \selectlanguage [0]{\@gobble}%
\providecommand \bibinfo  [0]{\@secondoftwo}%
\providecommand \bibfield  [0]{\@secondoftwo}%
\providecommand \translation [1]{[#1]}%
\providecommand \BibitemOpen [0]{}%
\providecommand \bibitemStop [0]{}%
\providecommand \bibitemNoStop [0]{.\EOS\space}%
\providecommand \EOS [0]{\spacefactor3000\relax}%
\providecommand \BibitemShut  [1]{\csname bibitem#1\endcsname}%
\let\auto@bib@innerbib\@empty
\bibitem [{\citenamefont {Von~Boehm}\ and\ \citenamefont
  {Bak}(1979)}]{von1979devil}%
  \BibitemOpen
  \bibfield  {author} {\bibinfo {author} {\bibfnamefont {J.}~\bibnamefont
  {Von~Boehm}}\ and\ \bibinfo {author} {\bibfnamefont {P.}~\bibnamefont
  {Bak}},\ }\bibfield  {title} {\bibinfo {title} {Devil's stairs and the
  commensurate-commensurate transitions in cesb},\ }\href@noop {} {\bibfield
  {journal} {\bibinfo  {journal} {Physical Review Letters}\ }\textbf {\bibinfo
  {volume} {42}},\ \bibinfo {pages} {122} (\bibinfo {year} {1979})}\BibitemShut
  {NoStop}%
\bibitem [{\citenamefont {Hasan}\ and\ \citenamefont
  {Kane}(2010)}]{hasan2010colloquium}%
  \BibitemOpen
  \bibfield  {author} {\bibinfo {author} {\bibfnamefont {M.~Z.}\ \bibnamefont
  {Hasan}}\ and\ \bibinfo {author} {\bibfnamefont {C.~L.}\ \bibnamefont
  {Kane}},\ }\bibfield  {title} {\bibinfo {title} {Colloquium: topological
  insulators},\ }\href@noop {} {\bibfield  {journal} {\bibinfo  {journal}
  {Reviews of modern physics}\ }\textbf {\bibinfo {volume} {82}},\ \bibinfo
  {pages} {3045} (\bibinfo {year} {2010})}\BibitemShut {NoStop}%
\bibitem [{\citenamefont {Dong}\ \emph {et~al.}(2015)\citenamefont {Dong},
  \citenamefont {Liu}, \citenamefont {Cheong},\ and\ \citenamefont
  {Ren}}]{dong2015multiferroic}%
  \BibitemOpen
  \bibfield  {author} {\bibinfo {author} {\bibfnamefont {S.}~\bibnamefont
  {Dong}}, \bibinfo {author} {\bibfnamefont {J.-M.}\ \bibnamefont {Liu}},
  \bibinfo {author} {\bibfnamefont {S.-W.}\ \bibnamefont {Cheong}},\ and\
  \bibinfo {author} {\bibfnamefont {Z.}~\bibnamefont {Ren}},\ }\bibfield
  {title} {\bibinfo {title} {Multiferroic materials and magnetoelectric
  physics: symmetry, entanglement, excitation, and topology},\ }\href@noop {}
  {\bibfield  {journal} {\bibinfo  {journal} {Advances in Physics}\ }\textbf
  {\bibinfo {volume} {64}},\ \bibinfo {pages} {519} (\bibinfo {year}
  {2015})}\BibitemShut {NoStop}%
\bibitem [{\citenamefont {Juraschek}\ \emph {et~al.}(2017)\citenamefont
  {Juraschek}, \citenamefont {Fechner}, \citenamefont {Balatsky},\ and\
  \citenamefont {Spaldin}}]{juraschek2017dynamical}%
  \BibitemOpen
  \bibfield  {author} {\bibinfo {author} {\bibfnamefont {D.~M.}\ \bibnamefont
  {Juraschek}}, \bibinfo {author} {\bibfnamefont {M.}~\bibnamefont {Fechner}},
  \bibinfo {author} {\bibfnamefont {A.~V.}\ \bibnamefont {Balatsky}},\ and\
  \bibinfo {author} {\bibfnamefont {N.~A.}\ \bibnamefont {Spaldin}},\
  }\bibfield  {title} {\bibinfo {title} {Dynamical multiferroicity},\
  }\href@noop {} {\bibfield  {journal} {\bibinfo  {journal} {Physical Review
  Materials}\ }\textbf {\bibinfo {volume} {1}},\ \bibinfo {pages} {014401}
  (\bibinfo {year} {2017})}\BibitemShut {NoStop}%
\bibitem [{\citenamefont {McIver}\ \emph {et~al.}(2020)\citenamefont {McIver},
  \citenamefont {Schulte}, \citenamefont {Stein}, \citenamefont {Matsuyama},
  \citenamefont {Jotzu}, \citenamefont {Meier},\ and\ \citenamefont
  {Cavalleri}}]{mciver2020light}%
  \BibitemOpen
  \bibfield  {author} {\bibinfo {author} {\bibfnamefont {J.~W.}\ \bibnamefont
  {McIver}}, \bibinfo {author} {\bibfnamefont {B.}~\bibnamefont {Schulte}},
  \bibinfo {author} {\bibfnamefont {F.-U.}\ \bibnamefont {Stein}}, \bibinfo
  {author} {\bibfnamefont {T.}~\bibnamefont {Matsuyama}}, \bibinfo {author}
  {\bibfnamefont {G.}~\bibnamefont {Jotzu}}, \bibinfo {author} {\bibfnamefont
  {G.}~\bibnamefont {Meier}},\ and\ \bibinfo {author} {\bibfnamefont
  {A.}~\bibnamefont {Cavalleri}},\ }\bibfield  {title} {\bibinfo {title}
  {Light-induced anomalous hall effect in graphene},\ }\href@noop {} {\bibfield
   {journal} {\bibinfo  {journal} {Nature physics}\ }\textbf {\bibinfo {volume}
  {16}},\ \bibinfo {pages} {38} (\bibinfo {year} {2020})}\BibitemShut {NoStop}%
\bibitem [{\citenamefont {Bloch}\ \emph {et~al.}(2022)\citenamefont {Bloch},
  \citenamefont {Cavalleri}, \citenamefont {Galitski}, \citenamefont {Hafezi},\
  and\ \citenamefont {Rubio}}]{Bloch2022}%
  \BibitemOpen
  \bibfield  {author} {\bibinfo {author} {\bibfnamefont {J.}~\bibnamefont
  {Bloch}}, \bibinfo {author} {\bibfnamefont {A.}~\bibnamefont {Cavalleri}},
  \bibinfo {author} {\bibfnamefont {V.}~\bibnamefont {Galitski}}, \bibinfo
  {author} {\bibfnamefont {M.}~\bibnamefont {Hafezi}},\ and\ \bibinfo {author}
  {\bibfnamefont {A.}~\bibnamefont {Rubio}},\ }\bibfield  {title} {\bibinfo
  {title} {Strongly correlated electron--photon systems},\ }\href@noop {}
  {\bibfield  {journal} {\bibinfo  {journal} {Nature}\ }\textbf {\bibinfo
  {volume} {606}},\ \bibinfo {pages} {41} (\bibinfo {year} {2022})}\BibitemShut
  {NoStop}%
\bibitem [{\citenamefont {Wessels}\ \emph {et~al.}(2016)\citenamefont
  {Wessels}, \citenamefont {Vogel}, \citenamefont {T{\"o}dt}, \citenamefont
  {Wieland}, \citenamefont {Meier},\ and\ \citenamefont
  {Drescher}}]{Wessels2016}%
  \BibitemOpen
  \bibfield  {author} {\bibinfo {author} {\bibfnamefont {P.}~\bibnamefont
  {Wessels}}, \bibinfo {author} {\bibfnamefont {A.}~\bibnamefont {Vogel}},
  \bibinfo {author} {\bibfnamefont {J.-N.}\ \bibnamefont {T{\"o}dt}}, \bibinfo
  {author} {\bibfnamefont {M.}~\bibnamefont {Wieland}}, \bibinfo {author}
  {\bibfnamefont {G.}~\bibnamefont {Meier}},\ and\ \bibinfo {author}
  {\bibfnamefont {M.}~\bibnamefont {Drescher}},\ }\bibfield  {title} {\bibinfo
  {title} {Direct observation of isolated damon-eshbach and backward volume
  spin-wave packets in ferromagnetic microstripes},\ }\href@noop {} {\bibfield
  {journal} {\bibinfo  {journal} {Scientific reports}\ }\textbf {\bibinfo
  {volume} {6}},\ \bibinfo {pages} {22117} (\bibinfo {year}
  {2016})}\BibitemShut {NoStop}%
\bibitem [{\citenamefont {Disa}\ \emph {et~al.}(2020)\citenamefont {Disa},
  \citenamefont {Fechner}, \citenamefont {Nova}, \citenamefont {Liu},
  \citenamefont {F{\"o}rst}, \citenamefont {Prabhakaran}, \citenamefont
  {Radaelli},\ and\ \citenamefont {Cavalleri}}]{Disa2020}%
  \BibitemOpen
  \bibfield  {author} {\bibinfo {author} {\bibfnamefont {A.~S.}\ \bibnamefont
  {Disa}}, \bibinfo {author} {\bibfnamefont {M.}~\bibnamefont {Fechner}},
  \bibinfo {author} {\bibfnamefont {T.~F.}\ \bibnamefont {Nova}}, \bibinfo
  {author} {\bibfnamefont {B.}~\bibnamefont {Liu}}, \bibinfo {author}
  {\bibfnamefont {M.}~\bibnamefont {F{\"o}rst}}, \bibinfo {author}
  {\bibfnamefont {D.}~\bibnamefont {Prabhakaran}}, \bibinfo {author}
  {\bibfnamefont {P.~G.}\ \bibnamefont {Radaelli}},\ and\ \bibinfo {author}
  {\bibfnamefont {A.}~\bibnamefont {Cavalleri}},\ }\bibfield  {title} {\bibinfo
  {title} {Polarizing an antiferromagnet by optical engineering of the crystal
  field},\ }\href@noop {} {\bibfield  {journal} {\bibinfo  {journal} {Nature
  Physics}\ }\textbf {\bibinfo {volume} {16}},\ \bibinfo {pages} {937}
  (\bibinfo {year} {2020})}\BibitemShut {NoStop}%
\bibitem [{\citenamefont {Yan}\ \emph {et~al.}(2009)\citenamefont {Yan},
  \citenamefont {Song}, \citenamefont {Mak}, \citenamefont {Chatzakis},
  \citenamefont {Maultzsch},\ and\ \citenamefont {Heinz}}]{yan2009time}%
  \BibitemOpen
  \bibfield  {author} {\bibinfo {author} {\bibfnamefont {H.}~\bibnamefont
  {Yan}}, \bibinfo {author} {\bibfnamefont {D.}~\bibnamefont {Song}}, \bibinfo
  {author} {\bibfnamefont {K.~F.}\ \bibnamefont {Mak}}, \bibinfo {author}
  {\bibfnamefont {I.}~\bibnamefont {Chatzakis}}, \bibinfo {author}
  {\bibfnamefont {J.}~\bibnamefont {Maultzsch}},\ and\ \bibinfo {author}
  {\bibfnamefont {T.~F.}\ \bibnamefont {Heinz}},\ }\bibfield  {title} {\bibinfo
  {title} {Time-resolved raman spectroscopy of optical phonons in graphite:
  Phonon anharmonic coupling and anomalous stiffening},\ }\href@noop {}
  {\bibfield  {journal} {\bibinfo  {journal} {Physical Review B}\ }\textbf
  {\bibinfo {volume} {80}},\ \bibinfo {pages} {121403} (\bibinfo {year}
  {2009})}\BibitemShut {NoStop}%
\bibitem [{\citenamefont {Kang}\ \emph {et~al.}(2010)\citenamefont {Kang},
  \citenamefont {Abdula}, \citenamefont {Cahill},\ and\ \citenamefont
  {Shim}}]{kang2010lifetimes}%
  \BibitemOpen
  \bibfield  {author} {\bibinfo {author} {\bibfnamefont {K.}~\bibnamefont
  {Kang}}, \bibinfo {author} {\bibfnamefont {D.}~\bibnamefont {Abdula}},
  \bibinfo {author} {\bibfnamefont {D.~G.}\ \bibnamefont {Cahill}},\ and\
  \bibinfo {author} {\bibfnamefont {M.}~\bibnamefont {Shim}},\ }\bibfield
  {title} {\bibinfo {title} {Lifetimes of optical phonons in graphene and
  graphite by time-resolved incoherent anti-stokes raman scattering},\
  }\href@noop {} {\bibfield  {journal} {\bibinfo  {journal} {Physical Review
  B}\ }\textbf {\bibinfo {volume} {81}},\ \bibinfo {pages} {165405} (\bibinfo
  {year} {2010})}\BibitemShut {NoStop}%
\bibitem [{\citenamefont {Malard}\ \emph {et~al.}(2009)\citenamefont {Malard},
  \citenamefont {Pimenta}, \citenamefont {Dresselhaus},\ and\ \citenamefont
  {Dresselhaus}}]{malard2009raman}%
  \BibitemOpen
  \bibfield  {author} {\bibinfo {author} {\bibfnamefont {L.}~\bibnamefont
  {Malard}}, \bibinfo {author} {\bibfnamefont {M.~A.}\ \bibnamefont {Pimenta}},
  \bibinfo {author} {\bibfnamefont {G.}~\bibnamefont {Dresselhaus}},\ and\
  \bibinfo {author} {\bibfnamefont {M.}~\bibnamefont {Dresselhaus}},\
  }\bibfield  {title} {\bibinfo {title} {Raman spectroscopy in graphene},\
  }\href@noop {} {\bibfield  {journal} {\bibinfo  {journal} {Physics reports}\
  }\textbf {\bibinfo {volume} {473}},\ \bibinfo {pages} {51} (\bibinfo {year}
  {2009})}\BibitemShut {NoStop}%
\bibitem [{\citenamefont {Nemanich}\ \emph {et~al.}(1977)\citenamefont
  {Nemanich}, \citenamefont {Lucovsky},\ and\ \citenamefont
  {Solin}}]{nemanich1977infrared}%
  \BibitemOpen
  \bibfield  {author} {\bibinfo {author} {\bibfnamefont {R.}~\bibnamefont
  {Nemanich}}, \bibinfo {author} {\bibfnamefont {G.}~\bibnamefont {Lucovsky}},\
  and\ \bibinfo {author} {\bibfnamefont {S.}~\bibnamefont {Solin}},\ }\bibfield
   {title} {\bibinfo {title} {Infrared active optical vibrations of graphite},\
  }\href@noop {} {\bibfield  {journal} {\bibinfo  {journal} {Solid State
  Communications}\ }\textbf {\bibinfo {volume} {23}},\ \bibinfo {pages} {117}
  (\bibinfo {year} {1977})}\BibitemShut {NoStop}%
\bibitem [{\citenamefont {Ferrari}\ and\ \citenamefont
  {Basko}(2013)}]{ferrari2013raman}%
  \BibitemOpen
  \bibfield  {author} {\bibinfo {author} {\bibfnamefont {A.~C.}\ \bibnamefont
  {Ferrari}}\ and\ \bibinfo {author} {\bibfnamefont {D.~M.}\ \bibnamefont
  {Basko}},\ }\bibfield  {title} {\bibinfo {title} {Raman spectroscopy as a
  versatile tool for studying the properties of graphene},\ }\href@noop {}
  {\bibfield  {journal} {\bibinfo  {journal} {Nature nanotechnology}\ }\textbf
  {\bibinfo {volume} {8}},\ \bibinfo {pages} {235} (\bibinfo {year}
  {2013})}\BibitemShut {NoStop}%
\bibitem [{\citenamefont {Davydov}(1964)}]{davydov1964theory}%
  \BibitemOpen
  \bibfield  {author} {\bibinfo {author} {\bibfnamefont {A.~S.}\ \bibnamefont
  {Davydov}},\ }\bibfield  {title} {\bibinfo {title} {The theory of molecular
  excitons},\ }\href@noop {} {\bibfield  {journal} {\bibinfo  {journal} {Soviet
  Physics Uspekhi}\ }\textbf {\bibinfo {volume} {7}},\ \bibinfo {pages} {145}
  (\bibinfo {year} {1964})}\BibitemShut {NoStop}%
\bibitem [{\citenamefont {F{\"o}rst}\ \emph {et~al.}(2011)\citenamefont
  {F{\"o}rst}, \citenamefont {Manzoni}, \citenamefont {Kaiser}, \citenamefont
  {Tomioka}, \citenamefont {Tokura}, \citenamefont {Merlin},\ and\
  \citenamefont {Cavalleri}}]{forst2011nonlinear}%
  \BibitemOpen
  \bibfield  {author} {\bibinfo {author} {\bibfnamefont {M.}~\bibnamefont
  {F{\"o}rst}}, \bibinfo {author} {\bibfnamefont {C.}~\bibnamefont {Manzoni}},
  \bibinfo {author} {\bibfnamefont {S.}~\bibnamefont {Kaiser}}, \bibinfo
  {author} {\bibfnamefont {Y.}~\bibnamefont {Tomioka}}, \bibinfo {author}
  {\bibfnamefont {Y.-n.}\ \bibnamefont {Tokura}}, \bibinfo {author}
  {\bibfnamefont {R.}~\bibnamefont {Merlin}},\ and\ \bibinfo {author}
  {\bibfnamefont {A.}~\bibnamefont {Cavalleri}},\ }\bibfield  {title} {\bibinfo
  {title} {Nonlinear phononics as an ultrafast route to lattice control},\
  }\href@noop {} {\bibfield  {journal} {\bibinfo  {journal} {Nature Physics}\
  }\textbf {\bibinfo {volume} {7}},\ \bibinfo {pages} {854} (\bibinfo {year}
  {2011})}\BibitemShut {NoStop}%
\bibitem [{\citenamefont {Landau}(1932)}]{landau1932theory}%
  \BibitemOpen
  \bibfield  {author} {\bibinfo {author} {\bibfnamefont {L.~D.}\ \bibnamefont
  {Landau}},\ }\bibfield  {title} {\bibinfo {title} {A theory of energy
  transfer ii},\ }\href@noop {} {\bibfield  {journal} {\bibinfo  {journal}
  {Phys. Z. Sowjetunion}\ }\textbf {\bibinfo {volume} {2}},\ \bibinfo {pages}
  {19} (\bibinfo {year} {1932})}\BibitemShut {NoStop}%
\bibitem [{\citenamefont {Zener}(1932)}]{zener1932non}%
  \BibitemOpen
  \bibfield  {author} {\bibinfo {author} {\bibfnamefont {C.}~\bibnamefont
  {Zener}},\ }\bibfield  {title} {\bibinfo {title} {Non-adiabatic crossing of
  energy levels},\ }\href@noop {} {\bibfield  {journal} {\bibinfo  {journal}
  {Proceedings of the Royal Society of London. Series A, Containing Papers of a
  Mathematical and Physical Character}\ }\textbf {\bibinfo {volume} {137}},\
  \bibinfo {pages} {696} (\bibinfo {year} {1932})}\BibitemShut {NoStop}%
\bibitem [{\citenamefont {Nicoletti}\ and\ \citenamefont
  {Cavalleri}(2016)}]{nicoletti2016nonlinear}%
  \BibitemOpen
  \bibfield  {author} {\bibinfo {author} {\bibfnamefont {D.}~\bibnamefont
  {Nicoletti}}\ and\ \bibinfo {author} {\bibfnamefont {A.}~\bibnamefont
  {Cavalleri}},\ }\bibfield  {title} {\bibinfo {title} {Nonlinear light--matter
  interaction at terahertz frequencies},\ }\href@noop {} {\bibfield  {journal}
  {\bibinfo  {journal} {Advances in Optics and Photonics}\ }\textbf {\bibinfo
  {volume} {8}},\ \bibinfo {pages} {401} (\bibinfo {year} {2016})}\BibitemShut
  {NoStop}%
\bibitem [{\citenamefont {Fano}(1961)}]{fano1961effects}%
  \BibitemOpen
  \bibfield  {author} {\bibinfo {author} {\bibfnamefont {U.}~\bibnamefont
  {Fano}},\ }\bibfield  {title} {\bibinfo {title} {Effects of configuration
  interaction on intensities and phase shifts},\ }\href@noop {} {\bibfield
  {journal} {\bibinfo  {journal} {Physical Review}\ }\textbf {\bibinfo {volume}
  {124}},\ \bibinfo {pages} {1866} (\bibinfo {year} {1961})}\BibitemShut
  {NoStop}%
\bibitem [{\citenamefont {Hasdeo}\ \emph {et~al.}(2014)\citenamefont {Hasdeo},
  \citenamefont {Nugraha}, \citenamefont {Dresselhaus},\ and\ \citenamefont
  {Saito}}]{hasdeo2014breit}%
  \BibitemOpen
  \bibfield  {author} {\bibinfo {author} {\bibfnamefont {E.~H.}\ \bibnamefont
  {Hasdeo}}, \bibinfo {author} {\bibfnamefont {A.~R.}\ \bibnamefont {Nugraha}},
  \bibinfo {author} {\bibfnamefont {M.~S.}\ \bibnamefont {Dresselhaus}},\ and\
  \bibinfo {author} {\bibfnamefont {R.}~\bibnamefont {Saito}},\ }\bibfield
  {title} {\bibinfo {title} {Breit-wigner-fano line shapes in raman spectra of
  graphene},\ }\href@noop {} {\bibfield  {journal} {\bibinfo  {journal}
  {Physical Review B}\ }\textbf {\bibinfo {volume} {90}},\ \bibinfo {pages}
  {245140} (\bibinfo {year} {2014})}\BibitemShut {NoStop}%
\bibitem [{\citenamefont {Zhang}\ \emph {et~al.}(2015)\citenamefont {Zhang},
  \citenamefont {Li}, \citenamefont {Xia}, \citenamefont {Liu}, \citenamefont
  {Shi}, \citenamefont {Luo}, \citenamefont {Hu}, \citenamefont {Richard},\
  and\ \citenamefont {Ding}}]{zhang2015observation}%
  \BibitemOpen
  \bibfield  {author} {\bibinfo {author} {\bibfnamefont {W.-L.}\ \bibnamefont
  {Zhang}}, \bibinfo {author} {\bibfnamefont {H.}~\bibnamefont {Li}}, \bibinfo
  {author} {\bibfnamefont {D.}~\bibnamefont {Xia}}, \bibinfo {author}
  {\bibfnamefont {H.}~\bibnamefont {Liu}}, \bibinfo {author} {\bibfnamefont
  {Y.-G.}\ \bibnamefont {Shi}}, \bibinfo {author} {\bibfnamefont
  {J.}~\bibnamefont {Luo}}, \bibinfo {author} {\bibfnamefont {J.}~\bibnamefont
  {Hu}}, \bibinfo {author} {\bibfnamefont {P.}~\bibnamefont {Richard}},\ and\
  \bibinfo {author} {\bibfnamefont {H.}~\bibnamefont {Ding}},\ }\bibfield
  {title} {\bibinfo {title} {Observation of a raman-active phonon with fano
  line shape in the quasi-one-dimensional superconductor k 2 cr 3 as 3},\
  }\href@noop {} {\bibfield  {journal} {\bibinfo  {journal} {Physical Review
  B}\ }\textbf {\bibinfo {volume} {92}},\ \bibinfo {pages} {060502} (\bibinfo
  {year} {2015})}\BibitemShut {NoStop}%
\bibitem [{\citenamefont {Bock}\ \emph {et~al.}(1999)\citenamefont {Bock},
  \citenamefont {Ostertun}, \citenamefont {Sharma}, \citenamefont
  {R{\"u}bhausen}, \citenamefont {Subke},\ and\ \citenamefont
  {Rieck}}]{bock1999anomalous}%
  \BibitemOpen
  \bibfield  {author} {\bibinfo {author} {\bibfnamefont {A.}~\bibnamefont
  {Bock}}, \bibinfo {author} {\bibfnamefont {S.}~\bibnamefont {Ostertun}},
  \bibinfo {author} {\bibfnamefont {R.~D.}\ \bibnamefont {Sharma}}, \bibinfo
  {author} {\bibfnamefont {M.}~\bibnamefont {R{\"u}bhausen}}, \bibinfo {author}
  {\bibfnamefont {K.-O.}\ \bibnamefont {Subke}},\ and\ \bibinfo {author}
  {\bibfnamefont {C.}~\bibnamefont {Rieck}},\ }\bibfield  {title} {\bibinfo
  {title} {Anomalous self-energy effects of the b 1 g phonon in y 1- x (pr, ca)
  x ba 2 cu 3 o 7 films},\ }\href@noop {} {\bibfield  {journal} {\bibinfo
  {journal} {Physical Review B}\ }\textbf {\bibinfo {volume} {60}},\ \bibinfo
  {pages} {3532} (\bibinfo {year} {1999})}\BibitemShut {NoStop}%
\end{thebibliography}%

\end{document}